\documentclass[onecolumn]{article}

\usepackage{graphicx}
\usepackage{amsmath}
\usepackage{amssymb}
\usepackage{bigstrut}
\usepackage{tabularx,ragged2e,booktabs,caption,authblk}
\newcolumntype{C}[1]{>{\Centering}m{#1}}

\everymath{\displaystyle}
\usepackage{fullpage}
\usepackage{cite}
\usepackage{float}

\begin{document}

\title{Re-Solving the Shepherding Problem: Lead When Possible, Herd When Necessary}

\author{Daniel Str\"{o}mbom\thanks{\mbox{Corresponding author: stroembp@lafayette.edu}}, Julianna Hoitt, Cameron Cloud.\\\small{Department of Biology, Lafayette College,  Easton, PA 18042, USA.}}

\date{}

\maketitle
\begingroup
\renewcommand\thefootnote{}
\footnotetext{This paper has been accepted for publication in the proceedings of ANTS~2026: 15th International Conference on Swarm Intelligence, to appear in Springer’s Lecture Notes in Computer Science (LNCS) series. Publication details (volume number, pages, DOI) will be added once available.}
\endgroup

\begin{abstract}
Designing systems for autonomous transport of groups of living agents has received a lot of attention in recent years due to a wealth of important potential applications. Biomimetic approaches are often sought, and a range of herding algorithms, inspired by how dogs herd sheep, as well as leadership algorithms mimicking leader-follower systems, have been introduced. However, they suffer from a common problem: shepherding algorithms require that agents evade the shepherd, and leading algorithms require that agents follow. This can cause problems in real-world applications where the behavioral responses of the agents to a transporter are likely to be heterogeneous over both long and short timescales. Here, we introduce an algorithm that adaptively switches between leading and herding depending on the response it receives from the agents to mitigate this problem. We show via simulation that this mixed algorithm can transport groups with any follower and evader composition, and we compare its performance with lead-only and herd-only algorithms. We also show that the mixed algorithm can deal with groups where individual agents randomly switch their strategy over time, as long as sufficient time is provided to complete the task relative to the switching rate. Given that our algorithm overcomes issues associated with herd-only and lead-only algorithms and might also, as a side effect, mitigate the issue of habituation to robotic transporters, it takes us one step closer to realizing many of the proposed applications for these types of algorithms.  
\end{abstract}


\section{Introduction}
Guiding or leading a group of individuals to a specific location is a task regularly performed by animals across taxa \cite{Sumpter2010,Ward2016}, ranging from primates leading troops to foraging sites \cite{Schaller1963} and sheepdogs herding flocks \cite{Coppinger2014,King2012}, to ants guiding or carrying conspecifics during nest relocation \cite{Holldobler1990,Dornhaus2004}. Understanding these phenomena is of interest not only in biology but also for engineering, as autonomous systems capable of transporting groups of inanimate objects or living agents have a wide range of potential applications \cite{King2023}, including wildlife conservation, environmental remediation, livestock management, crowd control, evacuation, and multi-robot coordination \cite{Fabregas2021,Fingas2001,Zahugi2013,Schultz1998,Li2022,EWT2002,Desholm2005,Brenner2005,Hughes2002,Hughes2005,Isobe2004,Turgut2008}.

Biomimetic approaches to collective transport are therefore common, with most recent work focusing on shepherding inspired by sheepdog behavior \cite{Long2020}. Numerous shepherding algorithms have been studied in simulation \cite{Li2022,Lien2004,Miki2006,Bennett2012,Strombom2014,Paranjape2018,Song2021,Auletta2022}, and several have been implemented on robotic platforms transporting inanimate or living agents \cite{Vaughan1998,Evered2014,Strombom2019}. Transport via leadership has also been explored \cite{Couzin2006}, and robotic leaders have been shown to elicit following behavior in a variety of animal systems \cite{Romano2018,Faria2010}. While effective in specific settings, both approaches share a fundamental limitation: herding requires agents to evade the transporter, whereas leading requires agents to follow it.

In natural groups, however, individuals often exhibit heterogeneous and context-dependent behavioral responses \cite{Jolles2020}, which may change over time due to habituation \cite{Evered2014} and can undermine the effectiveness of purely herd-based or leader-based strategies. Beyond these two modes, nature also exhibits mixed transport strategies that can be described as “lead when possible and force, carry, or herd when necessary,” as observed in some ant species during nest relocation \cite{Dornhaus2004}. This biomimetic alternative has not yet been considered in the context of autonomous collective transport of living agents, despite its potential relevance for many proposed applications \cite{King2023}.

At the same time, a growing body of work has begun to address heterogeneity explicitly, considering differences in responsiveness, social affiliations, or behavioral rule sets \cite{Bennett2021,Bennett2024,Himo2022,Fujioka2023,Hepworth2024}. These studies provide valuable insights into specific forms of heterogeneity but often rely on increasingly specialized control rules layered onto existing shepherding architectures. A complementary biomimetic perspective is that robust transport in nature frequently emerges from simple interaction mechanisms rather than progressively elaborate ones, a principle that has been explicitly exploited in earlier herding models \cite{Strombom2014}. Extending shepherding algorithms developed for different transport contexts \cite{Li2022,Lien2004,Song2021,Auletta2022} to incorporate alternative transport modes may therefore yield both more robust solutions and deeper theoretical insight into the autonomous transport of unwilling or variably responsive agents. 

\section{Model and Results}\label{sec2}
Here we introduce an algorithm for a transporter that adaptively switches between herding and leading depending on the response it receives from the agents. More specifically, we extend the shepherding model in \cite{Strombom2014} by adding a new mode of operation 'lead' that the transporter will autonomously and adaptively switch to from 'herding' if it detects that agents are following it. Another modification made here is that the (sheep-like) agents now adopt one of two strategies: 'follow' or 'evade' the transporter, and we introduce a parameter $p$ that represents the proportion of followers in the group of agents. We note that when $p=0$ (no followers) our new transporter algorithm is identical to the shepherding algorithm in \cite{Strombom2014}, when $p=1$ (all followers) the transporter will only lead, and for any $p$ in $(0,1)$ the transporter will employ the 'mixed' strategy where it leads if at least one agent is following and herds if no agent is following. We also introduce a parameter $\pi$ that represents the probability that an agent will switch strategy on each timestep.

\begin{figure}
\centering
\includegraphics[width=0.9\textwidth]{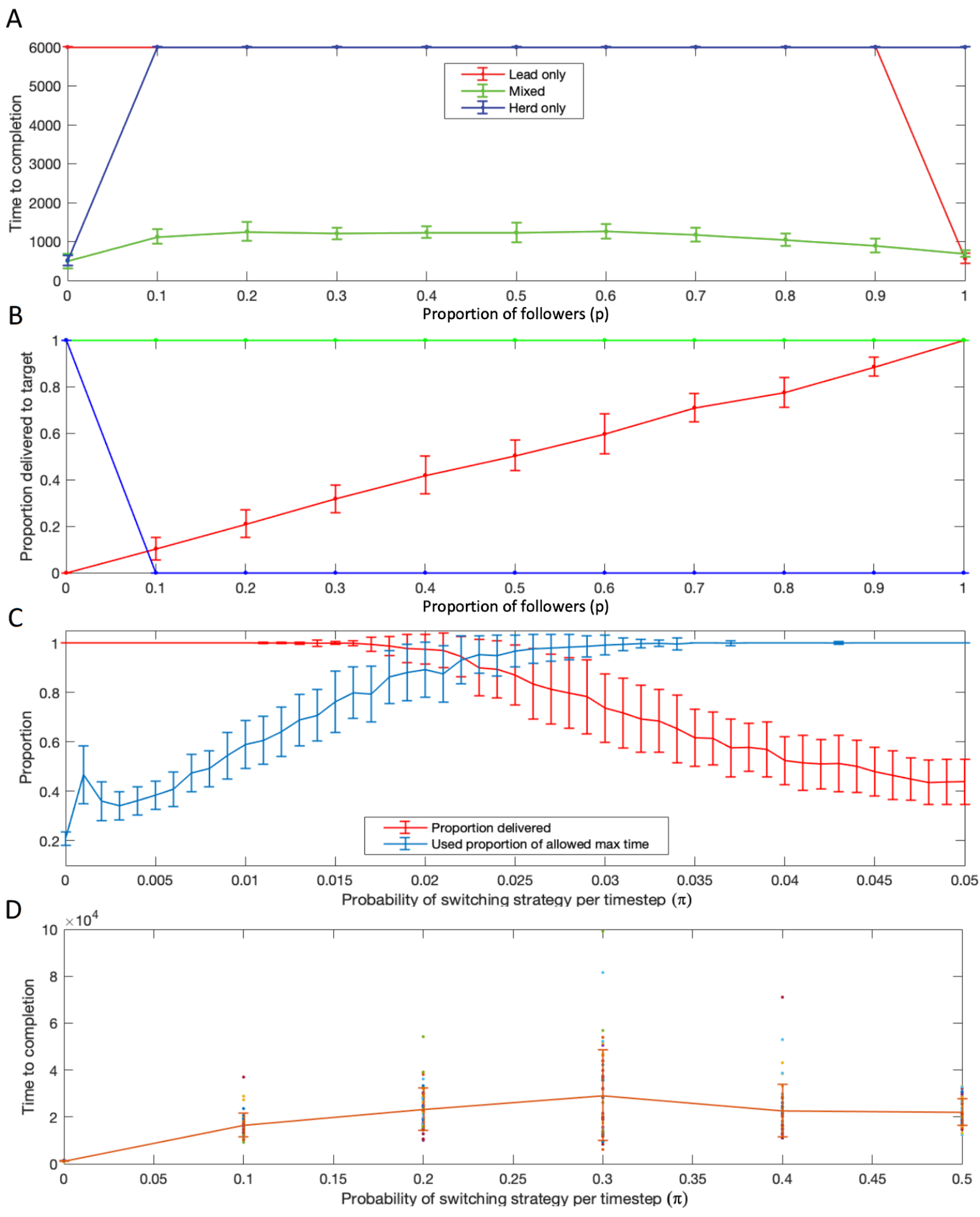}
\caption{Performance of the algorithms. (A)-(B) Comparison of the original herd-only algorithm, a lead-only algorithm, and the mixed 'lead when possible, herd when necessary' algorithm with respect to time to completion (A), and proportion of agents delivered to target (B), as a function of proportion of followers ($p$). (C) Performance of the mixed algorithm when agents switch strategy with probability $\pi$ per timestep. We see that for $\pi$ up to 0.01 the algorithm delivers all agents to the target within the allotted time, but as the rate of strategy switches increases the mixed algorithm's capacity to deliver the agents to the target within the allotted time decreases. (D) Time to completion for the mixed algorithm for $\pi$ from 0 to 0.5. The red curve indicates mean completion time, the bars represent the standard deviation, and the dots represent the actual completion times from each simulation. We note that the mean completion time for all $\pi$ is around or below 20000 timesteps and that overall only a few individual simulations that take far longer than that to complete.}
\label{fig:1}
\end{figure}

To study how the performance of the mixed, herd-only, and lead-only algorithms depends on the proportion of followers in a group, we ran simulations for values of $p$ from 0 to 1 and measured the time to completion and the proportion of agents delivered to the target. The mixed algorithm successfully delivers all agents for all proportions $p \in [0,1]$ within the allotted time (Fig.~\ref{fig:1}AB). In contrast, the herd-only algorithm succeeds only when all agents are evaders ($p=0$), since the presence of any followers prevents the transporter from maintaining an effective driving position. Conversely, the lead-only algorithm succeeds only when all agents are followers ($p=1$); if evaders are present, the transporter can guide the followers to the target but cannot induce the remaining agents to approach. The mixed algorithm avoids both failure modes by leading whenever at least one agent follows and reverting to herding otherwise, thereby guiding followers to the target before herding the remaining evaders.

To examine robustness to time-varying behavior, we next considered agents that stochastically switch between evading and following. For moderate switching rates (up to $\pi \approx 0.01$ per timestep), the mixed algorithm reliably delivers all agents within the fixed time horizon (Fig.~\ref{fig:1}C). As the switching rate increases further, the fixed time limit becomes insufficient and the success rate decreases. Removing the time constraint shows that the algorithm nevertheless succeeds even for high switching rates, including cases where agents switch strategy on average every other timestep (Fig.~\ref{fig:1}D).\\

See the Methods section for a more detailed description of the model and simulation protocols. 

\section{Discussion}\label{sec11}
Biomimetic solutions for autonomous transport of groups of inanimate and living agents are highly desirable because of the wealth of potential applications \cite{King2023}. Most proposed algorithmic solutions to these problems involve herding \cite{Long2020}, inspired by how sheepdogs herd sheep, and to a lesser degree leading, which is a ubiquitous phenomenon in social animal groups \cite{Sumpter2010,Ward2016}. However, both of these approaches suffer from a common issue that may render them ineffective for real-world applications: they are typically unable to deal with behavior-wise heterogeneous groups, in particular groups containing a mix of evaders and followers, or groups with individuals that may change behavioral strategies over time. Here we have extended a well-known shepherding algorithm to autonomously lead when possible and herd when necessary to mitigate this problem.

We establish that the mixed algorithm always succeeds regardless of the proportion of followers in a group if given sufficient time to complete the task (see Fig. \ref{fig:1}). Unlike herding and leading that only succeed in the extremes when either all are followers or all are evaders. For any given group composition, the mixed algorithm will attempt to move the agents towards the target in one of two ways and will proceed with whichever works at a given time, and because of this its success does not rely on agents being strictly evading or strictly following.

One practical issue with herding by robots is habituation, in particular, the agents may initially evade the robot but over time habituate and stop evading it \cite{Evered2014}, in which case a herd-only algorithm will stop working (see Fig. \ref{fig:1}). The mixed strategy may mitigate, or even exploit, habituation effects in transported animals. First, because it will switch between trying to lead and herd, if it receives ambiguous responses from the agents, which may make the transporter appear less predictable and therefore delay habituation. Secondly, if leading is attempted it may turn out that the agents can be led, in which case the more costly mode of transportation, herding, might be unnecessary. For animals, herding is also based on a fear response which may lead to stress \cite{Yaxley2021} and an increased risk of injury during transport. If the animals could instead be led, or persuaded to be led, this may improve animal welfare. In fact, it has been shown that some animals can be trained or persuaded to follow a robot \cite{Romano2018,Faria2010}, and perhaps the task of making animals follow an appropriate transporter is easier than coercing them to move via herding in some situations. We also note that there is anecdotal evidence that a sheepdog can lead a flock of sheep \cite{youtube} rather than herd them, and if the same dog could transport sheep using a combination of both these modes that would represent a direct real-world example of the mixed strategy. Finally, we note that the 'lead' mode in the algorithm does not depend on the real-world motivation of the agents for following the transporter. It can work equally well when animals are following the transporter because they perceive it as a leader or as a threat that they are trying to chase away. For example, the chasing and mobbing behavior seabirds have been observed to exhibit towards drones \cite{Frixione2021} may potentially be exploited to 'lead' them away. As with much prior work on autonomous collective transport, our results are obtained in simulation, which enables systematic exploration of heterogeneous and time-varying agent responses. We note, however, that closely related shepherding mechanisms have previously been implemented on simple robotic platforms and shown to robustly transport and reorganize objects under dynamic conditions when given sufficient time to complete the task \cite{Strombom2019}.

Given the shortcomings of herd-only and lead-only approaches, it is clear that new types of collective transport algorithms are required to achieve success with many of the proposed real-world applications for these systems, and to advance theoretical approaches beyond the current state of the art described, for example, in \cite{King2023,Papadopoulou2025}. In particular, algorithms that not only are robust to standard intrinsic or extrinsic noise, but also to substantial strategy differences in the individuals to be transported over both short and longer timescales given habituation \cite{Evered2014} and behavioral heterogeneity \cite{Jolles2020} issues. Recent work has begun to address specific forms of heterogeneity, for example, differences in social affiliations \cite{Bennett2021,Bennett2024}, responsiveness to the transporter \cite{Himo2022,Hepworth2024}, or behavioral rule sets \cite{Fujioka2023}, often by adding tailored control elements to existing shepherding architectures. These contributions are important, but often target particular scenarios and can increase control complexity. A complementary biomimetic perspective is that robust herding and guidance in nature often arises from simple interaction principles rather than from growing layers of specialized rules. Following this view, our mixed lead-herd approach introduces only a minimal switching mechanism informed by natural transport strategies, but nevertheless accommodates a fundamental axis of heterogeneity: whether individuals are attracted to or repelled from the transporter, potentially in a time-dependent manner. From a computational perspective, our adaptive switching mechanism introduces negligible overhead relative to standard shepherding algorithms, as it requires only monitoring the sign of the relative motion of the closest agent and does not scale with group size beyond computations already present in the baseline model. We see this as an initial step toward more general, biomimetic "third-wave" transport algorithms that remain effective across a wide range of heterogeneous behavioral responses, and that may complement and extend existing specialized approaches.

\section{Methods}
\subsection{Model}
Here we describe how we extend the shepherding model of \cite{Strombom2014} to implement a mixed “lead when possible, herd when necessary” transport strategy.

In the original model, $N$ sheep-like agents are herded by a single shepherd toward a target in an unbounded two-dimensional space. The position of the shepherd at time $t$ is denoted by $\bar{S}_t$, and the position of agent $i$ by $\bar{A}_{i,t}$. Agents farther than a distance $r_s$ from the shepherd perform small random movements, while agents within $r_s$ are repelled from the shepherd and attracted toward the center of mass of their $n$ nearest neighbors. Short-range agent-agent repulsion is also included. Under these interactions, the heading of agent $i$ at time $t+1$ is given by
\begin{equation}\label{eq:1}
\bar{H}_{i,t+1} = h \hat{H}_{i,t} + c\hat{C}_{i,t} + \rho_{a}\hat{R}_{i,t}^{a} + \rho_{s}\hat{R}_{i,t}^{s},
\end{equation}
where $\hat{H}_{i,t}$ is the normalized heading at time $t$, $\hat{C}_{i,t}$ is the direction toward the local center of mass, and $\hat{R}_{i,t}^{a}$ and $\hat{R}_{i,t}^{s}$ denote agent-agent and shepherd repulsion directions, respectively (see \cite{Strombom2014} for details).

Our first modification is to generalize the shepherd into a transporter, relative to which agents may adopt one of two behavioral strategies. Evading agents follow the dynamics in Eq. \ref{eq:1}, while following agents instead experience attraction toward the transporter, yielding
\begin{equation}\label{eq:2}
\bar{H}_{i,t+1} = h \hat{H}_{i,t} + c\hat{C}_{i,t} + \rho_{a}\hat{R}_{i,t}^{a} + \rho_{s}\hat{F}_{i,t}^{s},
\end{equation}
where $\hat{F}_{i,t}^{s} = -\hat{R}_{i,t}^{s}$. The two agent types therefore differ only in the sign of the transporter interaction.

Herding motion follows the standard collection-driving strategy of \cite{Strombom2014}: when the group is sufficiently cohesive, the transporter moves to a driving position behind the group relative to the target; otherwise, it collects the most distant agent by moving to a collection position. Our second modification is that the transporter dynamically switches from herding to leading when it detects that the closest agent is approaching rather than evading. In this case, the transporter adopts a heading directly toward the target and leads the following agent(s) until no agents continue to approach, at which point it resumes herding behavior. As in the original shepherding framework, the transporter is assumed to have access to the positions of all agents relative to itself and the target, with the additional ability to infer whether the closest agent is approaching or receding based on changes in relative distance over time; no explicit knowledge of agent strategies (follower or evader) or overall group composition is assumed.

\subsection{Simulation Protocols}
The target location is always located at the origin $(0,0)$ and the $N$ agents are initially assigned random positions in the upper right region of an $L \times L$ square and the transporter is released from near the target location in the lower left of this square. The heading update weights in Eqs. \ref{eq:1} and \ref{eq:2} are $h=0.5$, $\rho_{a}=2$, $\rho_{s}=1$, $c=1.5$ (Cf Table 1 in \cite{Strombom2014}) and we use $L=100$ and $N=46$ to match the simulation setup used to compare the original model to the empirical study \cite{King2012}. Using this common setup we ran three types of simulations (A), (B) and (C) here. 
(A) Simulations where the evader and follower status of each agent remains constant through the entire simulation to compare the herd-only, lead-only, and mixed algorithms. For each algorithm we ran 100 trials of up to 6000 timesteps for each proportion of followers $p$ from 0 to 1 in increments of 0.1 and measured the number of timesteps it took for the algorithm to complete the task (if it did, else 6000 was recorded) and the proportion of agents that was delivered to the target within the time limit. At the start of each simulation, each agent was randomly assigned to be a follower with probability $p$ (and thus an evader with probability $1-p$). The agents that were assigned to be evaders updated their headings using Eq. \ref{eq:1}, and those assigned to be followers updated their headings using Eq. \ref{eq:2}, throughout the entire simulation. The 6000-timestep limit was chosen based on pilot simulations to ensure completion well before this time. Simulations of this type were used to create Figs. \ref{fig:1}AB.
(B)-(C) Simulations where the evader and follower status of each agent (may) change throughout the simulation for the mixed algorithm only. At the start of each simulation, each agent is randomly assigned to be a follower with probability 0.5 so that on average half of the agents start off as followers and half evaders. Then, at each subsequent timestep each agent will switch strategy with probability $\pi$, where switching strategy corresponds to changing the update rule between Eqs. \ref{eq:1} and \ref{eq:2}. (B) Here we ran 100 simulations for each strategy switch rate $\pi$ from 0 to 0.05 in increments of 0.001 and measured task completion time and delivery success within 6000 timesteps. (C) Here we ran 100 simulations for each strategy switch rate $\pi$ from 0 to 0.5 in increments of 0.1, measuring completion time without a time limit (Figs. \ref{fig:1}CD).
 \bibliographystyle{unsrt} 





\end{document}